\documentclass{jfm}
\usepackage{graphicx}
\usepackage{newtxtext}
\usepackage{newtxmath}
\usepackage{natbib}
\usepackage{hyperref}
\usepackage{siunitx}
\usepackage{amsmath}
\usepackage{graphicx}

\hypersetup{
    colorlinks = true,
    urlcolor   = blue,
    citecolor  = black,
}

\newcommand{\RomanNumeralCaps}[1]
\linenumbers
\usepackage[dvipsnames]{xcolor}

\title{\textcolor{black}{Success and failure of the spreading law for large drops of dense granular suspensions}}

\author{Alice Pelosse
\corresp{\email{alice.pelosse@gmail.com}},
 {\'E}lisabeth Guazzelli
 \and Matthieu Roché}

\affiliation{Universit\'{e} Paris Cit\'{e}, CNRS,  Mati\`{e}re et Syst\`{e}mes Complexes UMR 7057, F-75013, Paris}

\begin{document}

\maketitle

\begin{abstract}
The spreading of \textcolor{black}{large viscous drops of density-matched suspensions of non-Brownian spheres on a smooth} solid surface is experimentally investigated at the global drop scale. 
\textcolor{black}{The focus is on
dense suspensions with a solid volume fraction equal to or greater than $40\%$ and for drops larger than the capillary length, i.e. for which the spreading is governed by the balance of  gravitational and viscous forces.} \textcolor{black}{Our findings indicate that all liquids exhibit a power law behaviour typical of gravity-driven dynamics, albeit with an effective suspension viscosity that is smaller than the bulk value.} 
When the height of the drop is of the order of the particle size, \textcolor{black}{the power law breaks down as the particles freeze while the contact line continues to advance.} 
\end{abstract}


\section{Introduction}
\label{sec:intro}

The spreading of a liquid on a solid substrate is a common phenomenon, e.g. \textcolor{black}{in coating processes of surfaces in industry. 
In many situations,} the spreading fluid is not always a pure liquid as it may carry rigid bodies, such as fine dust or solid particles added to the coating fluid. 
It is therefore of great importance to examine whether the spreading laws established for a pure fluid are still valid for such a complex fluid. 

In this work, we focus on granular suspensions, made of large particles (i.e. having diameter $d>$~\SI{10}{\micro\meter}) such that Brownian motion and colloidal forces are negligible.
Despite their high degree of heterogeneity at the microscopic scale, these particulate systems can be seen as continuous effective fluids at the macroscopic scale, with an effective bulk viscosity that solely depends on the particle volume fraction, $\phi$, and is notably independent of particle diameter $d$ for monodisperse rigid spheres \citep[see e.g.][]{GuazzelliPouliquen2018}.
This type of complex fluid has been extensively studied in bulk-flow situations.
\textcolor{black}{Granular suspensions confined by a free interface are less well documented, with interest in this area only recently emerging. 
The thinning of a fluid neck in the presence of particles has been reported in the situation of pinch-off of a suspension thread \citep{chateau2018pinch}, or during drop detachment \citep{furbank2004experimental, roche2011,bonnoit2012accelerated, thievenaz2022onset}.
In these configurations, a Newtonian regime is observed using the bulk viscosity of the suspension as long as the confinement on the particles is greater than a few particle diameters. 
As the pinch-off is approached, with only a few particles remaining trapped in the thinning filament, a second regime follows with accelerated dynamics compared to that expected from using the bulk viscosity of the suspension.
A similar scenario is observed during the break-up of a drop of suspension \citep{xu2023breakup}.
Another example of capillary flow occurs during dip coating, when a solid plate is pulled up from a bath of suspension.
In this configuration, a receding contact line is connected to the bath by a film of constant thickness determined by the fluid properties (viscosity, surface tension, and capillary length) and wall velocity \citep{derjaguin1943thickness,landau1988dragging}. 
Newtonian behaviour using the bulk viscosity of the suspension is also observed in the low confinement limit when the particles are smaller than the thickness of the entrained film  \citep{palma2019dip,gans2019dip}.
However, it is the viscosity of the suspending phase that must be accounted for when the film thickness is less than one particle diameter.
In all of these systems, a continuum approach utilising the suspension bulk viscosity can be employed, provided that the particle confinement is not too strong. However, such a macroscopic description fails when the size of the system approaches that of the particle.
The present paper examines a canonical capillary flow phenomenon, namely the spreading of a drop of suspension on a smooth solid substrate.
Previous studies have investigated the behaviour of a granular suspension in the vicinity of an advancing contact line \citep{zhao2020spreading,pelosse2023probing}.}
In that latter case, the relation between the dynamic contact angle and the \textcolor{black}{contact line} velocity is similar to the classic Cox-Voinov law observed in the case of a pure liquid \citep{voinov1976hydrodynamics, cox1986dynamics}.
However, the wetting viscosity involved in this law differs from that of the bulk as it depends not only on $\phi$ but also on $d$. 
This observation is linked to the ability of the particles to approach the contact line close enough to affect dissipation \textcolor{black}{locally} \citep{zhao2020spreading,pelosse2023probing}.

\textcolor{black}{The present work investigates the drop-scale spreading of viscous density-matched suspensions (having a volume $V_0$, a surface tension with air $\gamma$, and a density $\rho$) on a solid surface by recording the time evolution of their spreading radius, $R(t)$. 
For pure fluids, the radius growth is expected to follow \textcolor{black}{a series of power laws} \citep{tanner1979spreading, lopez1976spreading},
\begin{equation}
    \textcolor{black}{R(t)=A\,t^n},
    \label{eq:powerlaw}
\end{equation}
where both the factor $A$ and the exponent $n$ depend on the balance of stresses acting in the system. At short times, spreading is driven by the balance of capillary and viscous forces, leading to $A\propto(V_0^3\gamma/\eta)^{1/10}$ and $n=1/10$ \citep{de1985wetting}.
Conversely, at longer times, gravity prevails over capillarity and the spreading behaviour is established through the balance between gravity and viscous dissipation, resulting in $A\propto(V_0^3\rho g/\eta)^{1/8}$ and $n=1/8$ \citep{lopez1976spreading,hocking1983}.
The spreading law in the crossover regime between these two algebraic laws is indistinguishable from the gravity-driven power law \citep{brochard1991spreading,redon1992spreading}. Furthermore, this crossover occurs at shorter times as the drop volume increases \citep{cazabat1986dynamics,levinson1988spreading}.}

\textcolor{black}{This paper discusses the validity of the spreading power law \eqref{eq:powerlaw} for large drops of granular suspensions. For these large drops which can contain many particles, we observe a power-law growth with an exponent $n=1/8$. However, the effective viscosity inferred from the prefactor of the spreading law is smaller than the usual bulk viscosity of the suspension. At very long times and for large particles, the dynamics deviates from the power law and slows down. The experimental methods are described in detail in \S\,\ref{sec:methods}. The results of the spreading experiments are presented and discussed in \S\,\ref{sec:res}. Finally, concluding remarks are drawn in \S\,\ref{sec:conc}.} 

\section{Experimental methods}
\label{sec:methods}

Two different types of granular suspensions have been used in the experiments.
The first combination of particles and fluid consists of spherical polystyrene beads (Dynoseeds TS, Microbeads, Norway) suspended in a density-matched Newtonian PEG copolymer [Poly(ethylene glycol-ran-propylene glycol) monobutyl ether] (Sigma) widely used in previous experimental work \citep[see e.g.][]{boyer2011unifying, bougouin2017collapse}.
The fluid dynamic viscosity is measured to be $\eta_f=2.4\pm 0.1$~Pa.s at \SI{22}{\celsius} and its density $\rho=1056$~kg/m$^3$ is close to that of \textcolor{black}{the polystyrene spheres.
To evaluate the effect of density mismatch \textcolor{black}{over a time $\Delta t$}, we can estimate the typical drift for a single particle, $U\Delta t$, with $U = U_{Stokes}(1-\phi)^5= gd^2\Delta \rho(1-\phi)^5 /(18\eta)$.
In dense suspensions, the Stokes sedimentation speed, $U_{Stokes}$, must be corrected by a factor $(1-\phi)^5\simeq 0.08$ at $40 \%$ \citep{richardson1954sedimentation}.
This velocity estimate suggests that for a timescale $\Delta t = 10^3$~\si{s}, even a density mismatch of \SI{10}{\kilo\gram\per\cubic\metre} is negligible, as the drift of a single \SI{20}{\micro\metre}- or \SI{550}{\micro\metre}-particle is \SI{0.08}{\micro\metre} or \SI{57}{\micro\metre}, respectively.}
Different batches of particles are used with varying mean diameters $d$ = 20, 40, 80, 140, 250 and \SI{550}{\micro\meter} (with dispersion in size of 10\% or less). 
The suspension mixture is made by weighting a mass of suspending fluid and adding the amount of solid needed to reach the desired particle volume fraction $\phi = 40\,\%$ in most of the situations presented in \S\ref{sec:res}. 
Larger particle loadings are also investigated as described in \S\ref{sec:breakdown}. 
Mixing is achieved by first slowly stirring with a spatula and then using a rolling device overnight. 
The particles are found to be completely wet by the fluid and to experience no aggregation. 
The surface tension of the suspensions is that of the PEG copolymer, $\gamma=\gamma_f\simeq$~\SI{35}{\milli\newton\per\meter}, as confirmed by pendant drop experiments.
The second type of suspension consists of 60-\si{\micro\meter} PMMA spheres (Spheromers CA, Microbeads, Norway) immersed in a fluid chosen to match the density and index of the particles. 
This fluid is a mixture of Triton X-100 (73\,wt\%), zinc chloride (16\,wt\%), and water (11\,wt\%) having a dynamic viscosity $\eta_f=3.3\pm 0.1$~Pa.s at \SI{22}{\celsius}. 
A fluorescent dye, Rhodamine 6G, is added to the fluid to aid the visualisation with a laser sheet as described in \S\ref{sec:PIV}. 
Only a single  solid volume fraction of $\phi=40\%$ has been studied for this suspension mixture.

Drops comprising the smallest particles (from \SI{20}{\micro\meter} to \SI{250}{\micro\meter}) are made using a syringe pump pushing the target volume of fluid at a flow rate \SI{1}{\milli\liter\per\min} out of a \SI{3}{\milli\meter} needle (inner diameter). 
To avoid transient heterogeneity in the suspension microstructure, at least \SI{3}{\milli\liter} of suspension is discarded before any experiment. 
For the largest \SI{550}{\micro\meter} particles, due to concerns about confinement effects in the tubing, the drops are made manually with a spatula ensuring a better control of $\phi$.
In all cases, the drops are deposited onto (Neyco Fused Silica) \textcolor{black}{a quartz wafer} carefully enough to produce axisymmetric spreading.
To this extent, before each experiment, the substrate is cleaned thoroughly with deionised water and ethanol and treated with a plasma cleaner to avoid \textcolor{black}{pinning} of the advancing contact line. 
Under these conditions, the continuous phase of the suspensions wets completely the surface of the wafer.
For a given batch of same particle size, concentration, and drop volume, each experiment is repeated at least three times. Error bars in graphs are inferred from the standard deviation of the experimental measurements.

The spreading of the drop is recorded by a (Imaging Source) monochrome digital camera (with a spatial resolution of 23 pixels/mm) located \SI{50}{\centi\metre} above the solid substrate and operated at a frame rate of \SI{1}{\per\second} which provides a capture of both early and long-time dynamics. 
Automatic measurement of the radius is performed by a Python script that first blurs locally the raw picture over roughly 5\,pixels, then detects the edges of the drop using the (scikit-image toolbox) Canny edge detector with a Gaussian width of 1, and finally returns the best circle fitting the outer radius using the (scikit-image toolbox) Hough transform function. 
Additional side-view recording is performed by a camera positioned at a few centimetres from the drop (with spatial resolution of \SI{0.6}{\micro\meter}/pixel) and synchronised with the top-view camera. Post-treatment of the recording uses the Sobel filter to detect the drop contour and extract the drop profile as a function of time \citep{pelosse2023probing}. 

In our experiments, the initial time is ill-defined, see figure\,\ref{fig:rayon_vs_time}($a$).
Indeed, as the drop touches the solid surface and starts spreading, it is attached to the needle for a certain amount of time which goes from a few seconds to one minute depending on the drop \textcolor{black}{size and} composition. 
The growth of the radius is therefore only investigated once the drop is fully detached from the needle, i.e. at a delayed time $t_0$.
\textcolor{black}{Following previous experimental investigations of drop spreading \cite[see e.g.][]{redon1992spreading}, the} radius data are fitted with the function 
\begin{equation}
    R(t,t_0,A) = A (t+t_0)^{n}.
    \label{eq:shiftedPowerLaw}
\end{equation}
The fitted parameters $t_0$ and $A$ are estimated numerically. 
An alternative method is to use an initial radius instead of an initial time, i.e. $R(t,R_0,A^{\prime}) = R_0 + A^{\prime} t^{1/8}$ \citep{saiseau2022near}.
This is equally efficient\textcolor{black}{,} but we have chosen the time-offset correction in the present data analysis.
\textcolor{black}{As mentioned in \S\,\ref{sec:intro},} for smaller drops \textcolor{black}{made of pure silicone oils}, a transition from a (short-time) capillary to a (long-time) gravity regime has been reported \citep{cazabat1986dynamics, levinson1988spreading}.
\textcolor{black}{Such changes in power law are not observed for the large drop sizes investigated in the present work.}
It should be stressed that the time offset correction, \textcolor{black}{$t_0\simeq10$~s}, only affects the behaviour at \textcolor{black}{short times} and does not modify that at \textcolor{black}{long times,} which is driven by gravity with a power law exponent of 1/8 \textcolor{black}{for $t\gtrsim100$~s, as discussed in the following \S\,\ref{sec:Tannerlaw}.}

\section{Results}
\label{sec:res}

\subsection{\textcolor{black}{Radius growth}}
\label{sec:Tannerlaw}


\begin{figure}
    \centerline{\includegraphics[width = \linewidth]{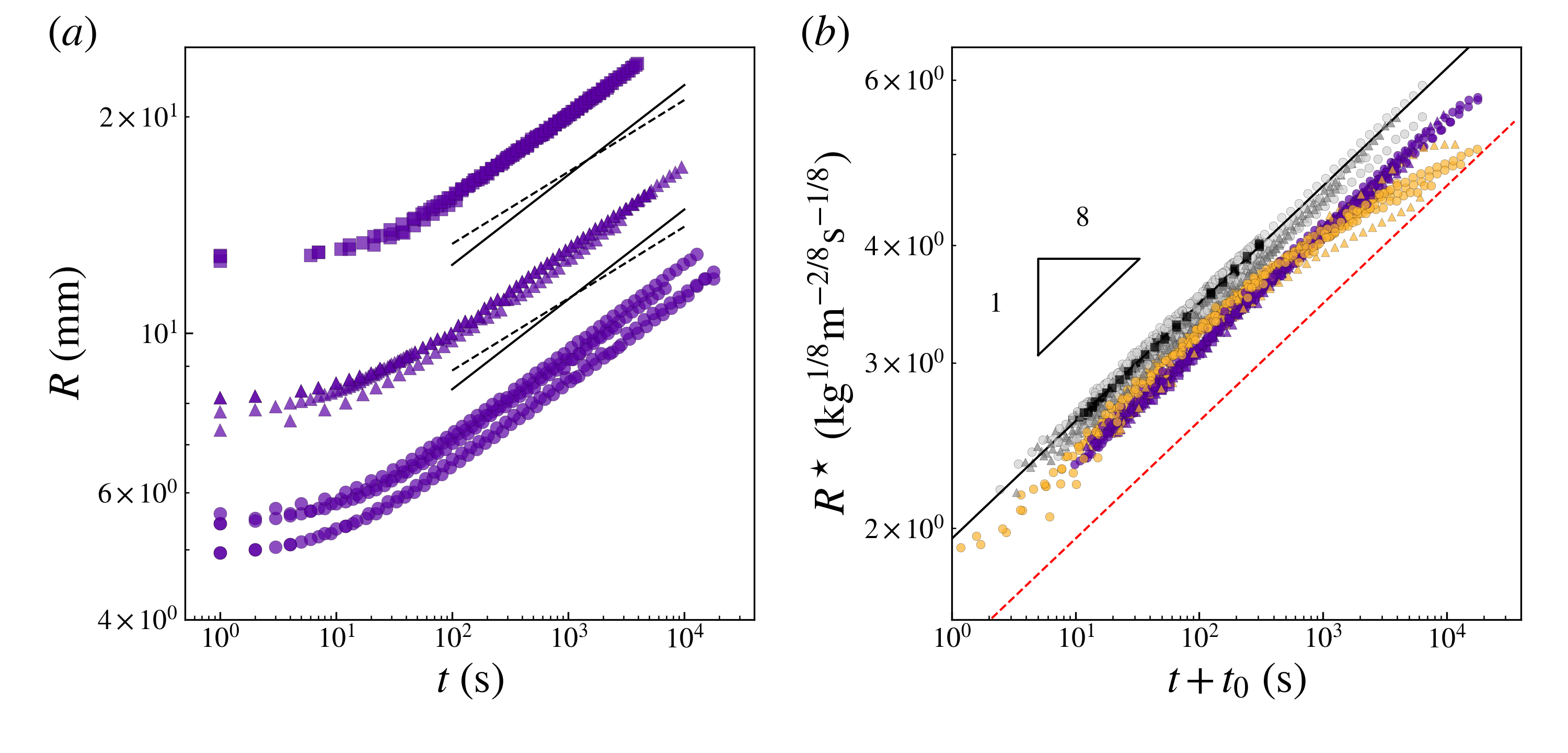}}
    \caption{\textcolor{black}{($a$) Raw radius $R$ versus time $t$ for drops made of a suspension of $d=$~40\,\si{\micro\meter} particles at $\phi=$~40\% with different drop volumes $V_0=$~\SI{100}{\micro\liter} ($\circ$ symbols), \SI{300}{\micro\liter} ($\triangle$ symbols), and \SI{1000}{\micro\liter} ($\square$ symbols). 
    Time $t=0$ is set when the drop detaches from the needle.
    Comparison with power laws with exponents $n=1/8$ (black solid lines) and $n=1/10$ (black dashed lines).
    ($b$) Normalised radius $R^\star=R(\eta_f/V_0^3)^{1/8}$ versus time (with the $t_0$ offset) for different drop volumes $V_0=$~100, 300, and \SI{1000}{\micro\liter} and different fluids: suspensions with $d=$~40\,\si{\micro\meter} (purple symbols) and 550\,\si{\micro\meter} (gold symbols) at $\phi=$~40\% and pure fluid (grey and black symbols). 
    Black solid line: best fit of the pure fluid data, $y = 1.94\,t^{1/8}$. 
    Red dashed line: expectation from suspension bulk viscosity $y = 1.94\,[t/\eta_s(40\%)]^{1/8} = 1.45\,t^{1/8}$.
    }}
\label{fig:rayon_vs_time}
\end{figure}

Figure~\ref{fig:rayon_vs_time} shows the typical time-evolution of the radius of drops consisting of pure fluid and granular suspensions.
In figure~\ref{fig:rayon_vs_time}($a$), the raw data are plotted versus time $t$ for suspensions made of 40-\si{\micro\meter} particles. The spreading at late times follows a gravity-driven dynamics, i.e. $R\sim t^{1/8}$.
Figure~\ref{fig:rayon_vs_time}($b$) displays the normalised radius $R^\star = R(\eta_f/V_0^3)^{1/8}$ versus $t+t_0$ for pure fluid and suspensions made of 40- and 550-\si{\micro\meter} particles \textcolor{black}{with $\eta_f$ the dynamic viscosity of the PEG copolymer measured by a capillary viscometer for each experiment.}
This radius normalisation removes volume and viscosity effects, as evidenced by the tight collapse of the curves for a given fluid, \textcolor{black}{but it should be stressed that this quantity possesses a dimension}.
The time offset corrects the initial bending seen in figure~\ref{fig:rayon_vs_time}($a$) as already explained at the end of \S\,\ref{sec:methods}. 
\textcolor{black}{This figure also displays the prediction for radius growth with the bulk viscosity of the suspensions. This trend fails at capturing any of our datasets, suggesting that the suspensions have an apparent viscosity during spreading that differs from its bulk value. We will address this observation in section \S \ref{sec:effective_viscosity}.}

\begin{figure}
    \centerline{\includegraphics[width =\linewidth]{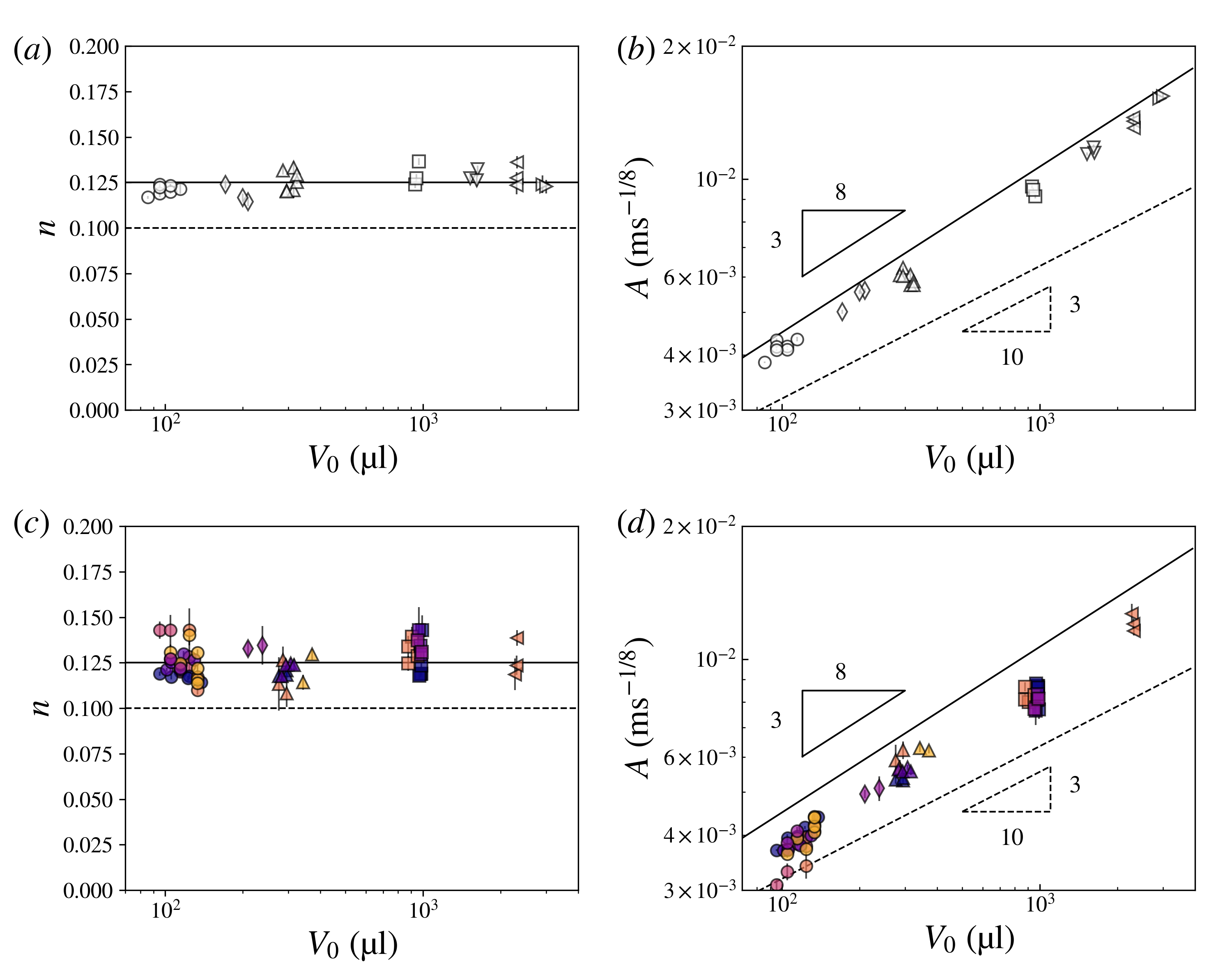}}
    \caption{\textcolor{black}{Spreading exponent $n$, and factor $A$, versus of drop volume $V_0$ for: ($a,b$) pure fluid, and ($c,d$) granular suspensions at $\phi=40\%$.
    Symbol shape corresponds to the drop volume: \SI{100}{\micro\liter} ($\circ$), \SI{300}{\micro\liter} ($\triangle$), and \SI{1000}{\micro\liter} ($\square$). 
    Symbol color correspond to the particle diameter from \SI{20}{\micro\meter} (blue) to \SI{550}{\micro\meter} (gold).
    Solid and dashed lines correspond to the theoretical predictions of the $V_0$-dependency in the gravity ($n=1/8, \,A\sim V_0^{3/8}$) and capillary ($n=1/10, \,A\sim V_0^{3/10}$) driven spreading, respectively.
    }}
\label{fig:fitting}
\end{figure}

\textcolor{black}{Figure~\ref{fig:fitting} compiles the values for the exponent $n$ and the prefactor $A$ obtained fromfitting power-law \eqref{eq:shiftedPowerLaw} to the raw data.}
\textcolor{black}{The exponents for pure fluid and granular suspensions are plotted in subplots $(a,c)$, respectively, and show no dependence on drop volume, $V_0$.
The values of the exponents are consistent with the conclusion drawn from figure 1, indicating that $n\simeq 1/8$. 
This is evidenced by the mean and standard deviation of $n$ being $(0.125,0.005)$ and $(0.125, 0.009)$ in the case of the pure fluid and the granular suspensions, respectively.}
\textcolor{black}{Furthermore, the confirmation of a gravity-driven spreading is also supported by the variations of $A$ with respect to the drop volume $V_0$. 
This is demonstrated by the relationship $A\sim V_0^{3/8}$, which is shown in figure~\ref{fig:fitting}($b,d$) for the pure fluid and the granular suspensions, respectively.}

\textcolor{black}{In summary, we observe that the radius of spreading of large drops of suspensions of any volume and concentration follows the power-law growth established for continuous Newtonian fluids in the gravity-driven regime.} 
However, there are two main differences  \textcolor{black}{evidenced in figure~\ref{fig:rayon_vs_time}($b$)}.  
First, the curves of the log-log plot for suspension drops exhibit a small vertical shift compared to those for the pure \textcolor{black}{fluid that} indicates that the factor $A$ of \textcolor{black}{the power} law differs when adding particles to the fluid, \textcolor{black}{but with a value that is not the one expected from the bulk suspension viscosity}. 
Second, for the largest particles ($d=$~\SI{550}{\micro\meter}), the dynamic slows down for $t\gtrsim$~\SI{200}{\second}. 
Deviations from \textcolor{black}{the power} law will be discussed in the following \S\,\ref{sec:effective_viscosity} and \ref{sec:breakdown}.
\subsection{\textcolor{black}{Effective viscosity}} 
\label{sec:effective_viscosity}
\begin{figure}
    \centerline{\includegraphics[width =\linewidth]{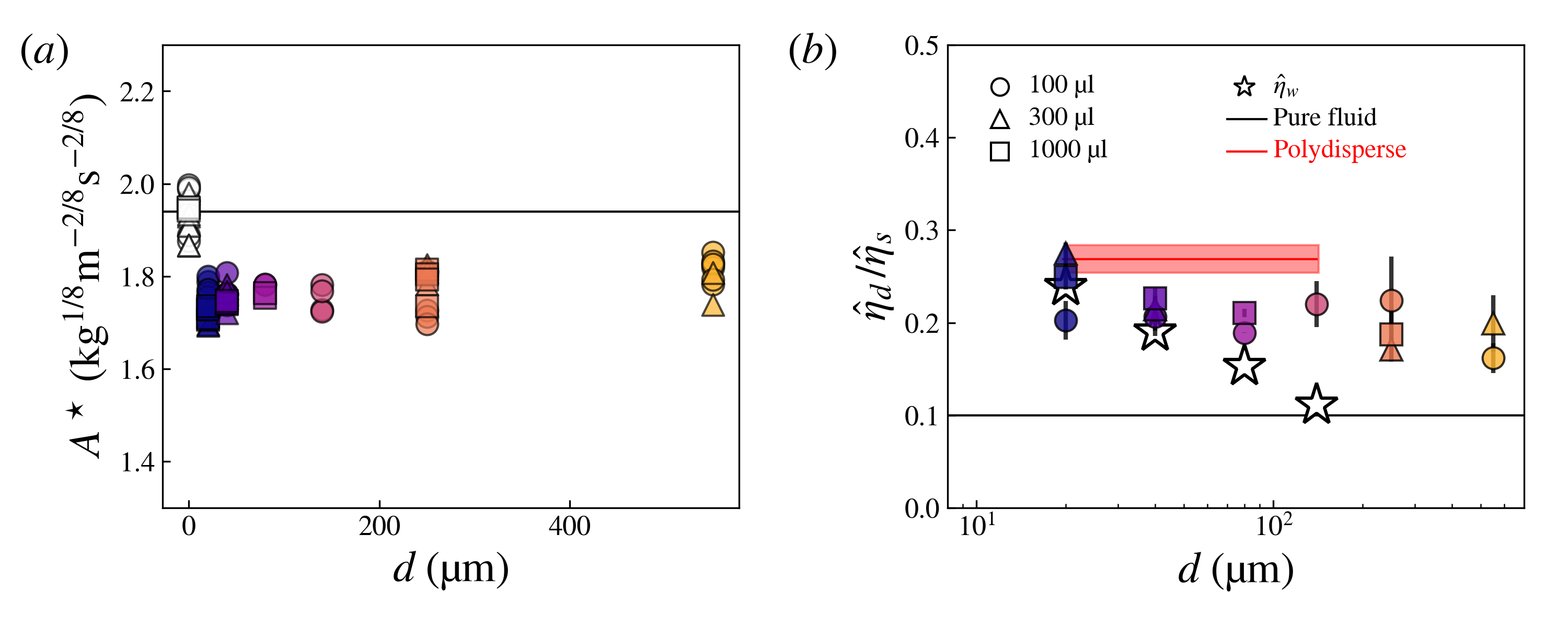}}
    \caption{\textcolor{black}{
    ($a$) Normalised factor $A^\star=A(\eta_f/V_0^3)^{1/8}$ versus particle diameter $d$ (from \SI{20}{\micro\meter} to \SI{550}{\micro\meter}) for drop volumes of \SIlist{100; 300; 1000}{\micro\liter}. 
    The white open symbols correspond to experiments with pure fluid. 
    ($b$) Ratio of the relative effective viscosity obtained from the power law, $\hat{\eta}_d$, to the relative bulk viscosity, $\hat{\eta}_s$, versus particle diameter $d$.
    The red horizontal line corresponds to this ratio for a polydisperse suspension made of particles of \SIlist{20;40;80;140}{\micro\meter}, each size representing 10~\% of the suspension volume.
    The star symbols correspond to the relative effective wetting viscosity of the same suspensions obtained from the Cox-Voinov law, $\hat{\eta}_w$ \citep{zhao2020spreading}.}}
\label{fig:viscosity}
\end{figure}
We start by discussing the impact of adding particles on the factor $A$.
In figure~\ref{fig:viscosity}($a$), the normalised factor, 
\begin{equation}
    A^\star=A \left(\frac{\eta_f}{V_0^3}\right)^{1/8},
\end{equation}
is plotted as a function of particle diameter, $d$, for all the experiments, \textcolor{black}{using the experimental drop volume, $V_0$, and the pure fluid dynamic viscosity, $\eta_f$ for the normalisation. 
The normalisation eliminates the $V_0$-dependence in $A \propto V_0^{3/8}$ found in  figure~\ref{fig:fitting}($b,d$)
as well as the effects of pure-fluid viscosity variations due to temperature or humidity}. 
\textcolor{black}{Note that this new quantity, $A^\star$ is not dimensionless.}
In figure~\ref{fig:viscosity}($a$), we see that the value of $A^\star$ is smaller for suspensions than for the pure continuous phase and its value is roughly independent of $d$.
However, the viscosity of the continuous phase $\eta_f$ used to compute $A^\star$ underestimates the \textcolor{black}{effective viscosity of the suspensions.}
Seeking to compare the effective viscosity derived from the power law with its bulk counterpart \textcolor{black}{$\hat{\eta}_s\cdot\eta_f$, with $\hat{\eta}_s$ the relative viscosity of the bulk suspension}, we write the suspension spreading viscosity of the drop as \textcolor{black}{$\hat{\eta}_d\cdot\eta_f$}.
\textcolor{black}{The non-dimensional quantity $\hat{\eta}_d$ is thus the relative effective viscosity during drop spreading and it quantifies the additional dissipation in the flow due to the addition of particles.}
To shift the suspension data onto the solid line corresponding to the pure fluid data in the figure\,\ref{fig:viscosity}($a$), the relative viscosity $\hat{\eta}_d$ should be computed as 
\begin{equation}
    \textcolor{black}{\hat{\eta}_d = \frac{\rho g}{\eta_f V_0^3}\left(\frac{k}{A}\right)^8=\left(\frac{A^\star_{f}}{A^\star}\right)^8,}
\end{equation}
where $k=0.61\pm 0.02$  has been inferred from the pure fluid data  (open symbols) using $A^\star_{f} = k(\rho g)^{1/8}=1.94$.
This relative viscosity extracted from \textcolor{black}{the power} law is reported in figure~\ref{fig:viscosity}($b$) for the different suspensions and drop volumes. It is found to be independent of $d$ and $V_0$ with a value $\hat{\eta}_d= 2.2 \pm 0.5$ \textcolor{black}{where the uncertainty is equal to the standard deviation.}

The \textcolor{black}{spreading} viscosity $\hat{\eta}_d$ is much smaller than the bulk viscosity of the suspension as the relative bulk viscosity is $\hat{\eta}_s \simeq 10$ at the same $\phi=40\%$. 
\textcolor{black}{This result is in line with the discrepancy observed in figure~\ref{fig:rayon_vs_time}($b$) between the experimental radius growth and the one anticipated using the suspension bulk viscosity shown by the red dashed line.
In other words, granular suspensions spread \textit{faster than would be expected from their bulk properties.}}

This relative \textcolor{black}{drop spreading} viscosity $\hat{\eta}_d$ which describes the global dynamics of the drop differs from the relative wetting viscosity \textcolor{black}{of the contact line} $\hat{\eta}_w$ extracted from the Cox-Voinov law found in the viscous–capillary region corresponding to the very close vicinity of the contact line, see the star symbols in figure~\ref{fig:viscosity}($b$).
In contrast with the constant behaviour of $\hat{\eta}_d$, $\hat{\eta}_w$ decreases with increasing $d$ and reaches a value of one (that corresponding to the pure fluid) for a cut-off size ($\approx$\,\SI{100}{\micro\meter}) above which particles are too large to affect dissipation in this region close to the contact line \citep{zhao2020spreading,pelosse2023probing}. 
This difference between $\hat{\eta}_w$ and $\hat{\eta}_d$ may not come as a surprise as it derives from two different energy balances at the local and global scales, i.e. from a balance of capillary and viscous forces at the local scale of the contact line and of gravity and viscous forces at the global scale of the drop.

\textcolor{black}{This decrease in viscosity, and therefore, in dissipation,} may be due to particle slip along the very smooth substrate.
Another difference may come from the suspension \textcolor{black}{microstructure, which} certainly is dissimilar in the present non-viscometric flow and in a pure shearing flow. 
Besides, confinement by both the substrate and the mobile interface may create layering that could span over 10 particle diameters at $\phi=40\%$ \citep{gallier2016effect} and end up in a strong dip in dissipation \citep{ramaswamy2017confinement}. 
\textcolor{black}{
However, experiments with polydisperse suspensions do not support this explanation. 
We use mixtures of 20, 40, 80, and 140~\si{\micro\meter} particles with a total solid volume fraction $\phi=40\,\%$, equal volume fractions of each particle diameter, and drop volumes of \SIlist{100;300;1000}{\micro\liter}.
It is anticipated that such polydispersity in the suspension will impede crystallisation.
If the decrease in dissipation observed in monomodal suspensions, $\hat{\eta}_d^{mono}\simeq0.22\hat{\eta}_s$, were due to crystallisation, the relative viscosity, $\hat{\eta}_d^{poly}$, of these polydisperse suspensions should be much closer to the relative bulk viscosity, $\hat{\eta}_s$.
However, the polymodal effective relative viscosity does not show any significant increase in dissipation as $\hat{\eta}_d^{poly}\simeq0.27 \hat{\eta}_s$, see the red line in  figure \ref{fig:viscosity}($b$).}


To  discriminate between possible causes of this smaller value of $\hat{\eta}_d$, we examine in detail the particulate flow during spreading in the following \S\,\ref{sec:PIV}.

\subsection{Visualisation of the particulate flow in the \textcolor{black}{one-dimensional} configuration}
\label{sec:PIV}

\begin{figure}
    \centerline{\includegraphics[width =\linewidth,trim= 0 8mm 0 8mm,clip]{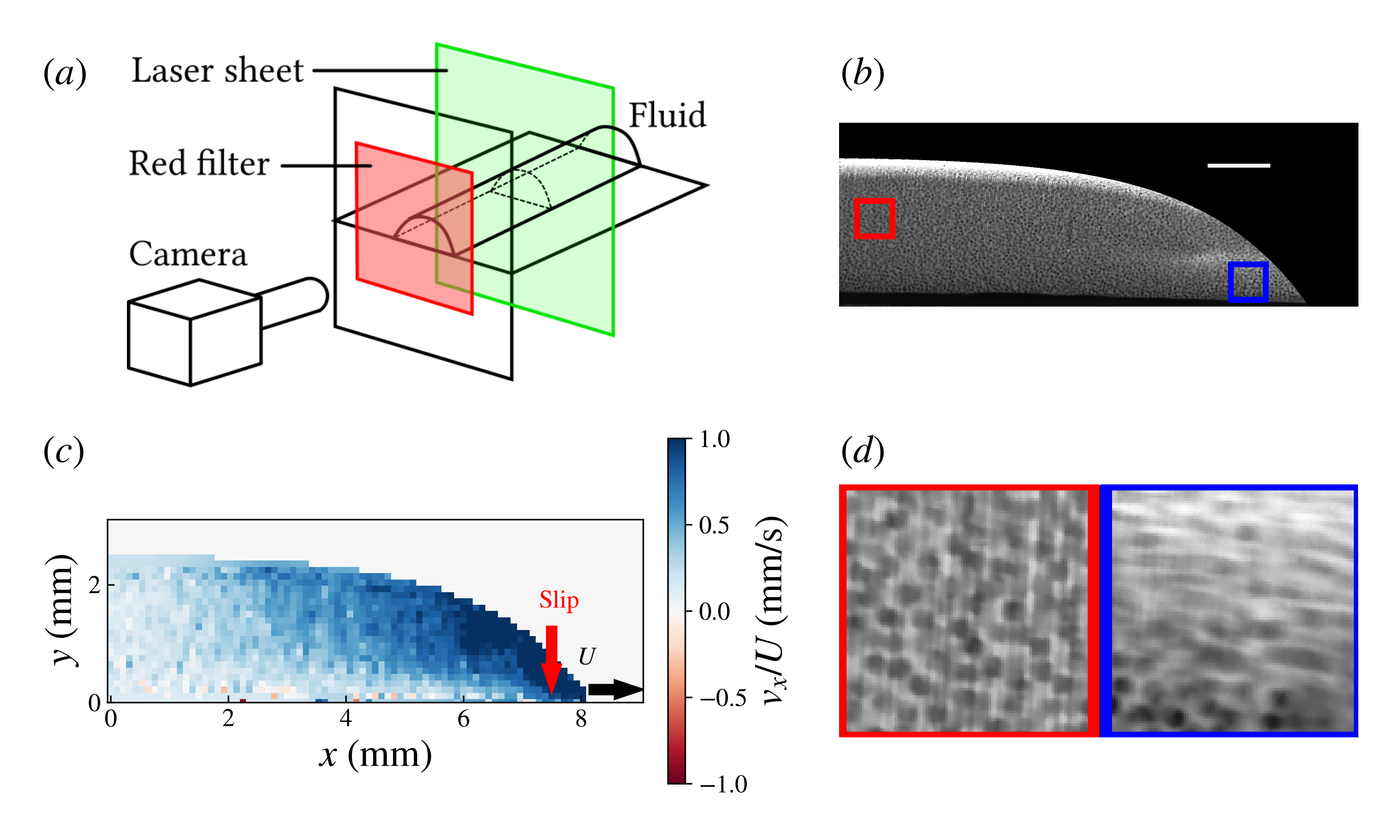}}
    \caption{($a$) Sketch of the apparatus used to visualize the particle flow in a 1D spreading with a laser sheet. A fluorescent dye suspended in the fluid is excited by the green light and fluoresces red light collected by a side view camera.
    ($b$) Imaging for a suspension made of 60-\si{\micro\meter} particles at $\phi=40\%$
    (scale bar: \SI{1}{\milli\meter}). See accompanying movie 1 in the supplementary material.
    ($c$) Horizontal speed $v_x$ from PIV analysis of the particle flow in ($b$).
    The speed is normalised by the contact line velocity $U$. \textcolor{black}{The red arrow indicates the occurrence of slip.}
    ($d$) Sum of raw pictures between $t$ and $t+\Delta t$ with $\Delta t = 2d/U$ in the corresponding red (left) and blue (right) squares in subplot $(b)$.  }
\label{fig:PIV}
\end{figure}

To gain further insight into particle motion during spreading, particle imaging is performed in a more convenient configuration, i.e. for spreading along only one direction, using a density and index-matched suspension described in  \S\,\ref{sec:methods}.
As sketched in figure~\ref{fig:PIV}($a$), imaging without optical distortion is achieved using two orthogonal microscope slides.
\textcolor{black}{On the horizontal slide, a 2-\si{\centi\meter} long thread of liquid is deposited with a spatula so that one end touches the vertical wall.
Care is taken to ensure a uniform thickness.}
A green laser sheet (Coherent lasers, $\lambda=$ \SI{532}{\nano\meter}, \SI{50}{\milli\watt}) illuminates the fluid through the bottom slide and a side view camera images the flow. 
\textcolor{black}{The distance between the vertical slide and the laser sheet is \SI{5}{\milli\meter}.}
The laser sheet is not deflected across the flat walls nor in the index-matched suspension and induces fluorescence of Rhodamine 6G in the illuminated slice (about \SI{30}{\micro\meter} thick) which is excited by the green light and emits red wavelengths.
This light is filtered by a high-pass red filter (Thorlabs, $\lambda>$~\SI{550}{\nano\meter}) and collected by the monochrome camera (Imaging Source, DMK33UX174, 2.3~MP) mounted with a zero distortion lens (Opto Engineering, MC100X), and orthogonal to the laser sheet. The spatial resolution is of \SI{3}{\micro\meter}/pixel and the acquisition frame-rate of \SI{1}{\per\second}.
With this apparatus, the non-fluorescing particles appear dark in the illuminated slice as seen in figure~\ref{fig:PIV}($b$) and the accompanying movie.
In figure~\ref{fig:PIV}($c$), the horizontal speed \textcolor{black}{ of the particulate phase is}  coming from the Particle Image Velocimetry (PIV) analysis performed with the openPIV module of Python on two consecutive frames.

The algorithm uses an interrogation window size of 16 pixels, a search area size of 18 pixels, and an overlap of 9 pixels. 
\textcolor{black}{The interrogation box typically contains 7-8 particles.}
The velocity is normalised by the speed of the advancing contact line (black arrow).
In figure~\ref{fig:PIV}($d$), the sum of the pictures between $t$ and $t+\Delta t$ is shown for $\Delta t =2d/U$ in two specific regions.
With this time interval, particles moving at a speed close to that of the contact line should not appear as fixed dark dots but should be blurred. 

Visualisation of the particulate flow during the \textcolor{black}{one-dimensional} spreading of the suspension unveils several significant elements.
Importantly, figure~\ref{fig:PIV}($b$) and its accompanying movie do not exhibit any clear development of layering or ordering during spreading, meaning that particle self-organisation cannot explain the strong decrease in viscosity as that reported by e.g.\,\cite{ramaswamy2017confinement}.
However, PIV measurements presented in figure~\ref{fig:PIV}($c$) and the sum of consecutive pictures in figure~\ref{fig:PIV}($d$) reveal two crucial features of the particle flow.
First, a strong particle slip develops along the wall near the advancing contact line as shown by the velocity signal in figure~\ref{fig:PIV}($c$) or by the smearing of dark dots near the edge in figure~\ref{fig:PIV}($d$) (right).
Second, in the centre, particles barely move as illustrated by the weak velocity signal from the PIV in figure~\ref{fig:PIV}($c$) or by the sum of the pictures in figure~\ref{fig:PIV}($d$) (left).
In this later figure, sharp dots at the centre (i.e. on the left) indicate that the particles have not moved within the time interval $\Delta t = 2d/U$, in contrast with the observations near the contact line (i.e. on the right) where the particulate phase undergoes a strong flow.

These observations of the structure of the particulate flow can explain the low value of the effective spreading viscosity displayed in figure\,\ref{fig:viscosity}($b$).
In rheometers, slip velocity of the particles leads to a smaller stress on the particles and results in an effective smaller viscosity \citep{jana1995apparent,yoshimura1988wall,ahuja2009slip}.
In spreading experiments, particles close to the tip exhibit a strong slip at roughly the contact line speed.
This slip is not surprising under such strong confinement and given the very smooth surface of the solid substrate.
Yet, slip by itself cannot account for such a large decrease in effective viscosity \textcolor{black}{and should also vary with particle size. 
These last two points suggest that at least one other mechanism contributes to the reduction in dissipation in drop spreading flows.} 
\textcolor{black}{In particular, this decrease could also be attributed to the near-frozen particulate phase at the centre which is such that the network of frictional particles in contact (as the particles are nearly jammed) acts more like a porous medium than like a flowing suspension in this region.
Interestingly, the effective viscosity of a porous media \citep{brinkman1949calculation, kim1985modelling,breugem2007effective} and similarly of a granular sediment bed \citep{vowinckel_rheology_2021} is found to be independent of particle size and to possess a value close to the Einstein viscosity ($\hat{\eta}_s = 1+2.5 \phi  =2$ at $\phi=0.4$), i.e. a value much smaller than that of a sheared dense suspension due to the screening of the long-range hydrodynamic interactions.}
This bears a striking resemblance to the behaviour of $\hat{\eta}_d$ in figure~\ref{fig:viscosity}($b$).

In summary, slip at the tip of the drop and porous-like behaviour at the centre of the drop could both account for the observed small value of the effective relative viscosity \textcolor{black}{of drop spreading}, $\hat{\eta}_d$.
\textcolor{black}{However, it should be noted that this analogy with a granular fixed bed to account for the decrease in dissipation seems less relevant in the case of the smallest particles, for which particle rearrangements can be observed in most of the drop until the end of the experiment.}

\subsection{Long time behaviour: validity of \textcolor{black}{the power} law}
\label{sec:breakdown}

\begin{figure}
    \centerline{
    \includegraphics[width =\linewidth,trim= 0 7mm 0 7mm,clip]{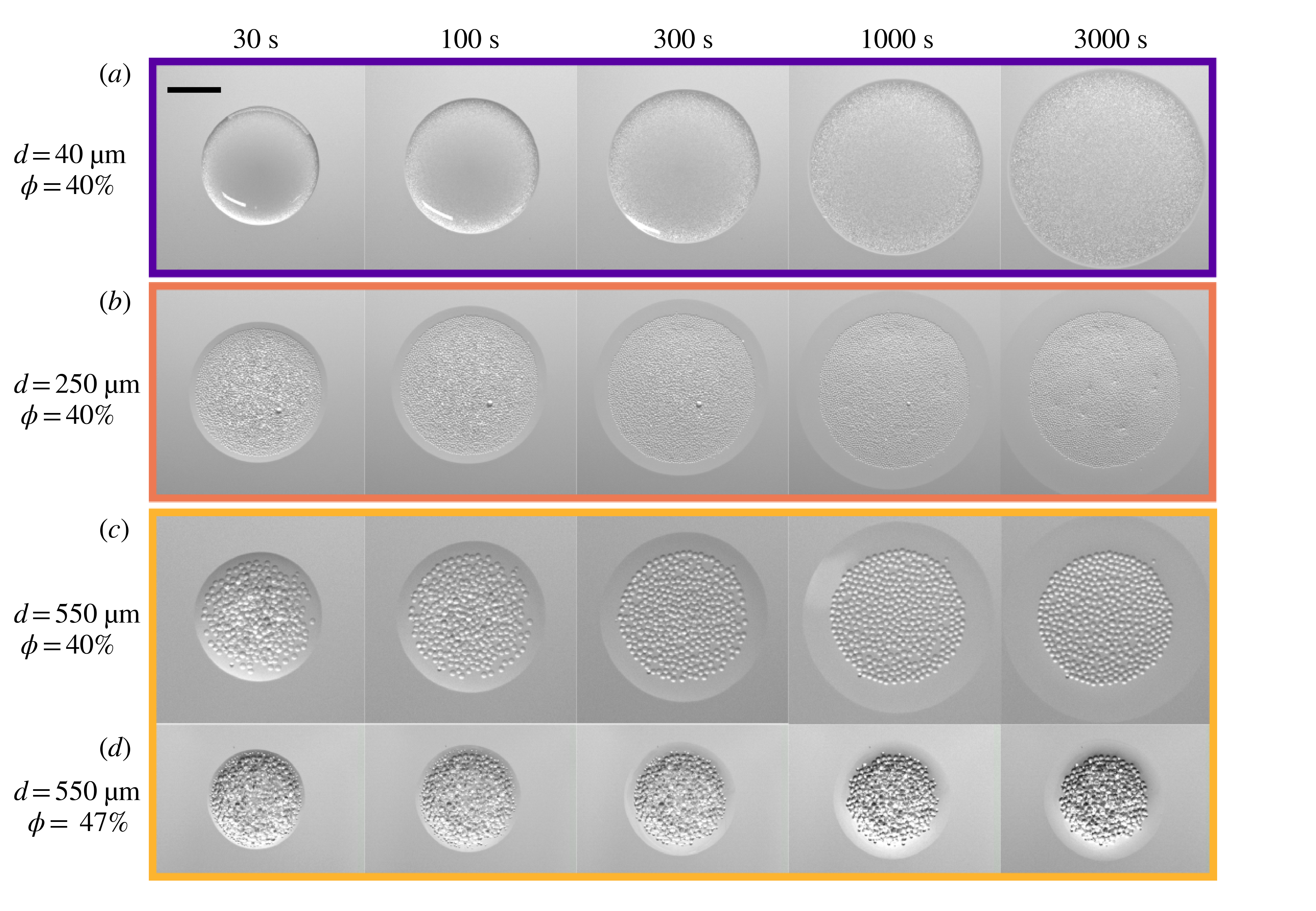}}
    \caption{Top view pictures of the spreading of different suspensions. From top to bottom: 
    ($a$) $d=$~\SI{20}{\micro\meter}, $\phi = 40\%$; 
    ($b$) $d=$~\SI{250}{\micro\meter}, $\phi = 40\%$; 
    ($c$) $d=$~\SI{550}{\micro\meter}, $\phi = 40\%$;
    ($d$) $d=$~\SI{550}{\micro\meter}, $\phi = 47\%$
    (scale bar: \SI{5}{\milli\meter}). See accompanying movie 2 in the supplementary material.}
\label{fig:top-view}
\end{figure}

We now turn to the slowdown of the spreading dynamics observed above \SI{200}{\second} for suspensions made of the \textcolor{black}{largest} \num{550}-\si{\micro\meter} particles. 
In figure~\ref{fig:top-view}, top view pictures of drop spreading evidence different patterns for suspensions having different particle size of \SIlist{40; 250; 550}{\micro\meter} at $\phi=40\%$ \textcolor{black}{in rows ($a$), ($b$), and ($c$), respectively}.
For the largest particles, top-views at a higher volume fraction, namely $\phi=47\%$, are also presented on the last row ($d$).
Clearly, the particle phase initially spreads with the fluid but then does not expand further beyond a critical radius which decreases with increasing particle size, as depicted on the first three rows ($a$), ($b$), ($c$), and with increasing particle volume fraction, as shown on the two last rows ($c$), ($d$). 
There is a striking difference between the behaviour for the smallest particles where the particle arrest is barely seen (top row) and that for the largest particles (bottom row) where particle freeze happens at \textcolor{black}{a very early stage}.
These observations suggest that the departure from \textcolor{black}{the power law can be associated with the freezing of the particle matrix while the contact line still moves.}

\begin{figure}
    \centerline{\includegraphics[width =\linewidth,trim= 0 7mm 0 6mm,clip]{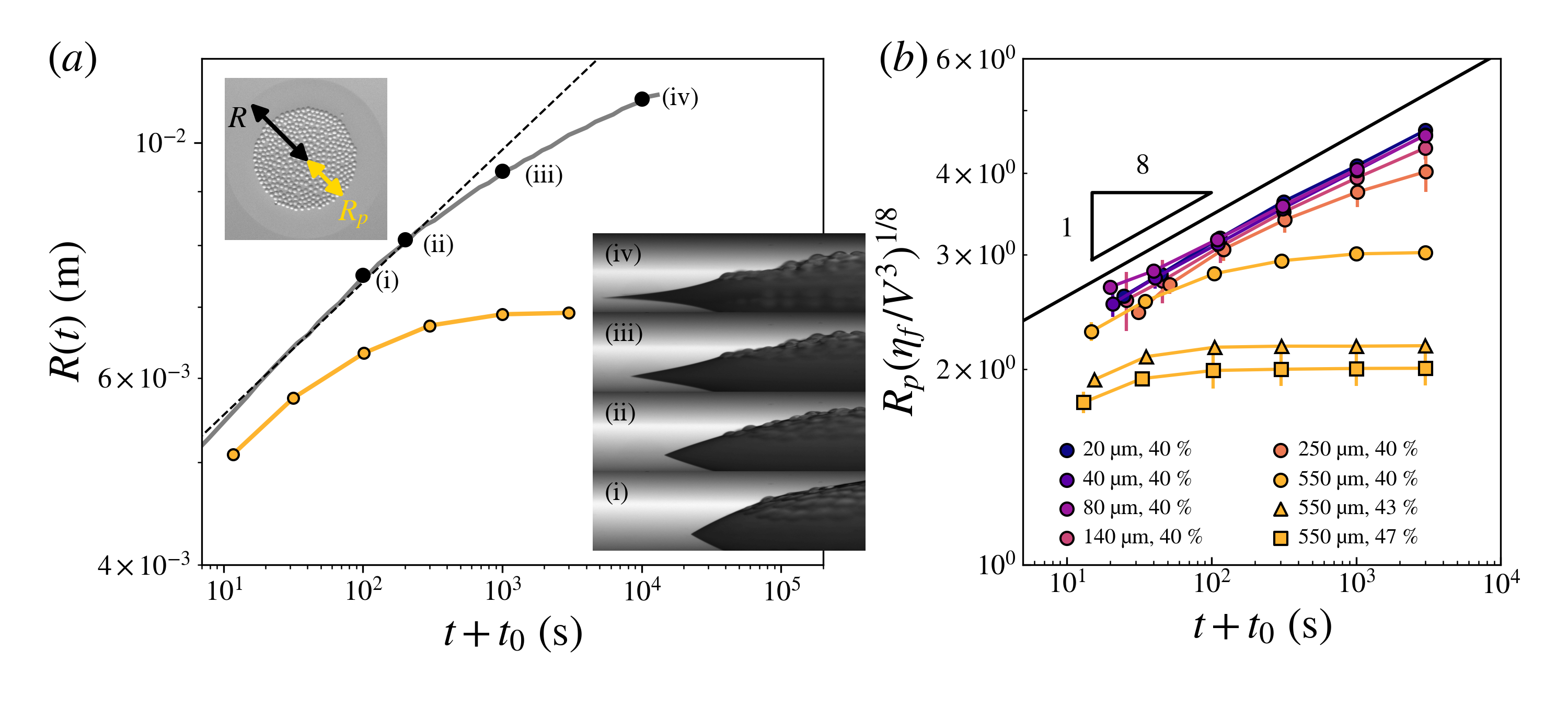}}
    \caption{($a$) Drop of suspension consisting of 550-\si{\micro\meter} particles with $V_0=$~\SI{100}{\micro\liter}. 
    Main graph: fluid radius $R$ (black) and particle radius $R_p$ (yellow) as a function of time. 
    The dashed line corresponds to \textcolor{black}{the power law}.
    Thumbnails (i) to (iv): side-view pictures corresponding to the black dots in the main graph.
    Top inset: top view picture with fluid ($R$) and particle ($R_p$) radii. 
    ($b$) Particle radius $R_p$ as a function of time for different suspensions (see legend) with a drop volume $V_0\simeq$~\SI{100}{\micro\liter}. 
    Error bars are computed from standard deviations over different experiments.}
\label{fig:breakdown}
\end{figure}

In figure~\ref{fig:breakdown}($a$), the fluid radius $R$ of a $\SI{100}{\micro\liter}$ drop of a suspension consisting of \SI{550}{\micro\meter} particles \textcolor{black}{at $\phi=40\%$} is plotted as a function of time (black solid line).
The particle spreading radius, $R_p$, defined in the top left inset, is also plotted with a yellow solid line.
Side-view snapshots numbered from (i) to (iv) taken during this spreading illustrate two different regimes.
First, for pictures (i) and (ii), the \textcolor{black}{slope of the drop profile is decreasing with increasing distance to the contact line} $R_p$ increases, meaning that particles move with the fluid, in agreement with the early time dynamics in figure~\ref{fig:top-view}. 
In this regime, the drop radius growth obeys \textcolor{black}{the power} law \textcolor{black}{with an exponent $n=1/8$,} and the suspension behaves like a continuous fluid.
Second, while the contact line still progresses from (ii) to (iv), $R_p$ saturates, i.e. the particles do not move and remain at the centre of the drop, also in agreement with observations from figure~\ref{fig:top-view}.
This translates into a significant slowdown of the drop radius growth $R$ and departure from \textcolor{black}{the power} law (black solid line). 
In this second regime, the continuous approach fails and the fluid drains out of the porous matrix consisting of the large particles.
\textcolor{black}{Then the slope of the drop profile increases with increasing distance to the edge, as seen in the thumbnails (iii) and (iv).} 
In these pictures, the particle protrusions become more significant and one can \textcolor{black}{identify in the picture (iv), away from the contact line, a region of constant height strongly distorting the free surface, which corresponds to} a monolayer of particles.

\textcolor{black}{Time evolutions of normalised particle radii for drops consisting of different particle sizes and volume fractions are plotted in figure~\ref{fig:breakdown}($b$)}. 
The same normalisation to remove volume and viscosity effects is used and the drop volumes are kept similar at approximately \SI{100}{\micro\liter}.
This graph confirms that particle freezing happens earlier with large particles and large volume fractions.
Conversely, for particle diameters $\lesssim$\SI{100}{\micro\meter}, $R_p$ follows a power law growth with an exponent $1/8$, i.e. the same dynamics as that predicted for a continuous fluid \textcolor{black}{for gravity-driven spreading}.

\begin{figure}
    \centerline{\includegraphics[width =1\linewidth,trim= 0 5mm 0 0,clip]{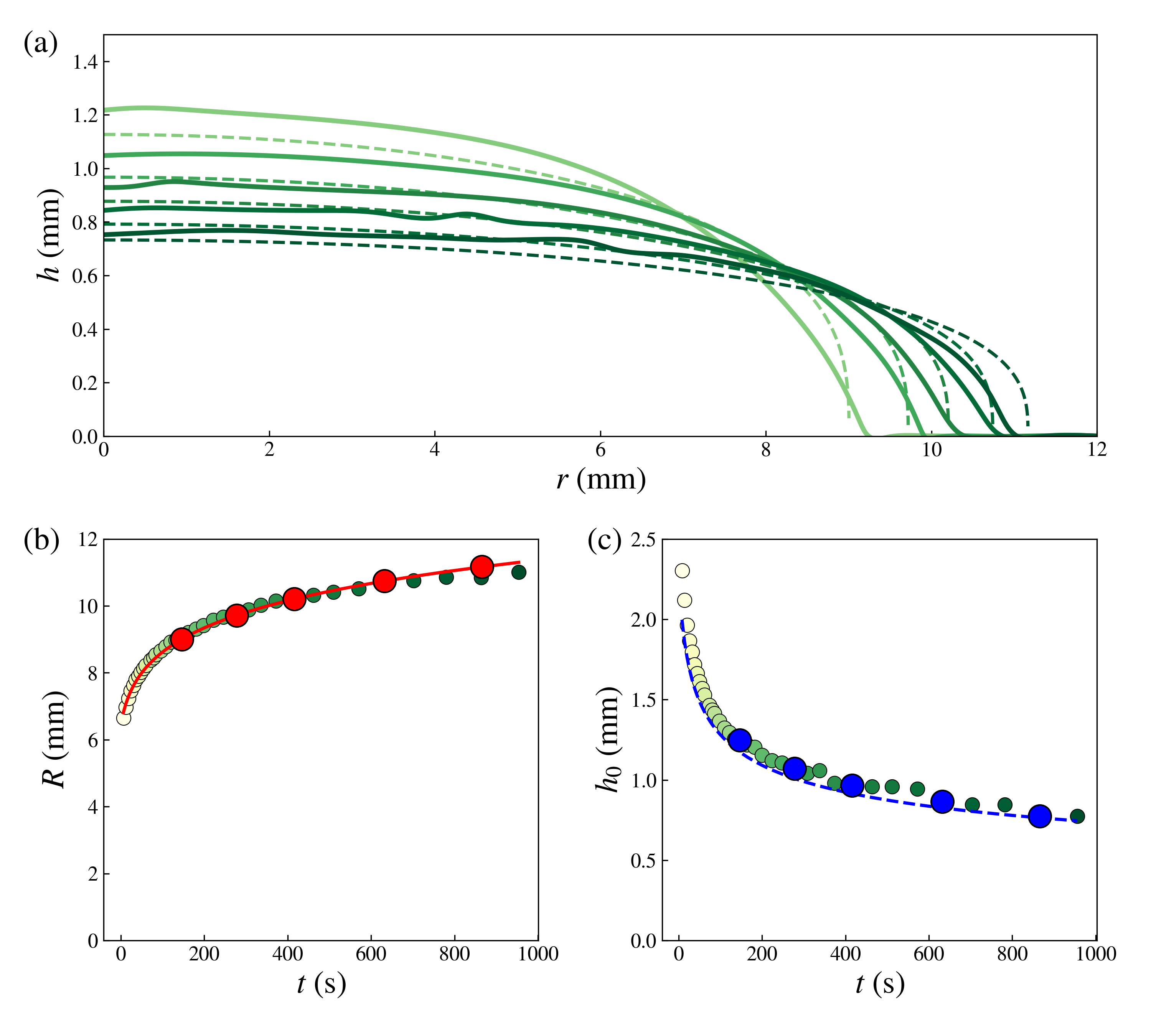}}
    \caption{\textcolor{black}{Spreading of a drop of suspension made of a \num{140}-\si{\micro\meter} particles with $V_0=$~\SI{215}{\micro\liter} and $\phi=40\%$.
    ($a$) Experimental profiles from side views (solid lines) and theoretical shapes. 
    Dashed lines: prediction from \eqref{eq:sol_lopez} given by \citet{lopez1976spreading}.
    ($b$) Dots: experimental radius $R$ as a function of time, red line: power-law growth $R = A\left(t+t_0\right)^{1/8}$, with the fitted parameters $A=$~4.79~10$^{-3}$~\si{\meter\second}$^{-1/8}$, $t_0=$~\SI{10.8}{\second}. 
    ($c$) Central thickness $h_0$ as a function of time\textcolor{black}{. 
    Dots: measurements from side views, dashed line:} prediction of \eqref{eq:sol_lopez} \textcolor{black}{using $R(t)$ fitted in subplot ($b$)}.
    The colour gradation indicates the time progression and the large dots in ($b$) and ($c$) correspond to the moments of the profiles plotted in ($a$).}
    }
\label{fig:profiles}
\end{figure}

To gain insight on the transition between these two regimes, the shape and thickness of the central part of the drop have been investigated. 
\textcolor{black}{For gravity-driven spreading, the theoretical prediction is given by \citet{lopez1976spreading},
\begin{equation}
    h(r,t)=\frac{4V_0}{3\pi R^2(t)}\left[1-\left(\frac{r}{R(t)}\right)^2\right]^{1/3},
    \label{eq:sol_lopez}
\end{equation}
where $r$ is the radial distance from the drop centre. 
Figure~\ref{fig:profiles}($a$) shows \textcolor{black}{at different times the profiles extracted from side views (solid lines) of a spreading droplet of suspensions made of 140-\si{\micro\meter} particles at $\phi=40~\%$ and with a volume $V_0=215$~\si{\micro\liter}.
The profiles at the center} of the droplet agree well with the theoretical prediction \eqref{eq:sol_lopez} (dashed lines), especially at long times. 
This agreement suggests that side-views are not required and drop profiles in the central region can be inferred using the drop radius $R(t)$ from top-view pictures.
The reader should note that the prediction \eqref{eq:sol_lopez} conflicts with experimental data at the droplet edge. 
This discrepancy was already highlighted by \citet{hocking1983} and arises from the fact that the central region of the droplet is not matched to the region near the contact line where capillary forces must be accounted for}.

In figure~\ref{fig:profiles}($b$), the radius data (computed from top views) are fitted with \textcolor{black}{the power} law (red solid line) and used in figure~\ref{fig:profiles}($c$) to compute the drop central thickness, $h_0$, according to \eqref{eq:sol_lopez} (blue dashed line). 
\textcolor{black}{In figure~\ref{fig:profiles}($c$), measurement of $h_0$ from side views shows excellent agreement \textcolor{black}{with the theoretical prediction from $R(t)$ data}.}



\begin{figure}
    \centerline{\includegraphics[width =0.7\linewidth]{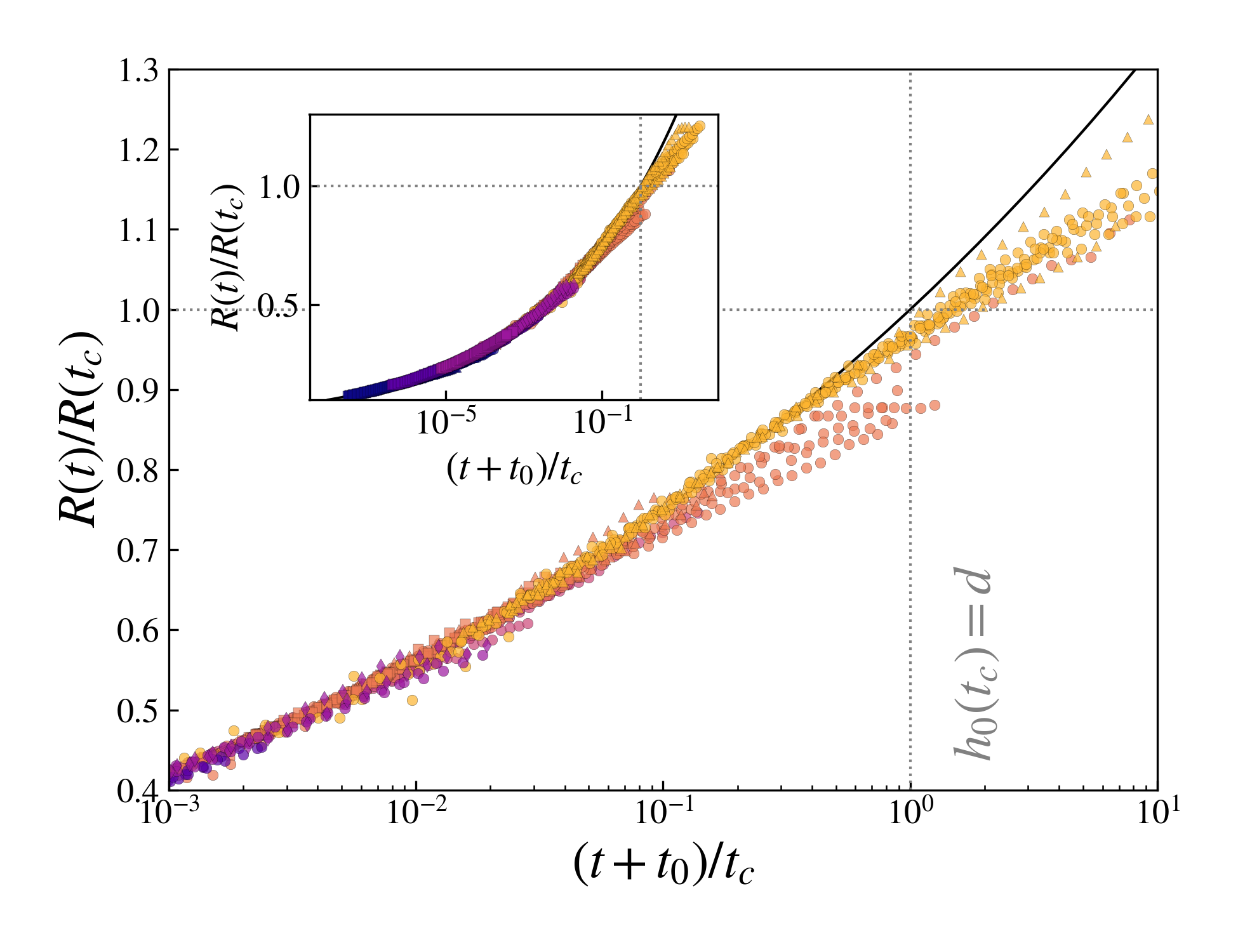}}
    \caption{\textcolor{black}{Radius as a function of time, normalised by $R_c$ and $t_c$, respectively, with $R_c=At_c^{1/8}$ and $h_0(R_c)=d$ (see legend in figure~\ref{fig:viscosity} for the symbols).
    Inset: Full-time and radius range,
    main graph: blow-up exhibiting the slowing dynamics.}}
\label{fig:hocking}
\end{figure}

The simplest criterion that can be proposed is that the breakdown of \textcolor{black}{the power-law radius growth} for granular suspensions happens when the drop thickness reaches approximately a particle diameter, as suggested by the side-view thumbnails in figure~\ref{fig:breakdown}($a$) where drop thickness is roughly 2-diameter thick at the transition for the 550-\si{\micro\meter} particles.
This can be estimated by evaluating the time $t_c$ (or equivalently the drop radius $R_c$) for which the central thickness $h_0(R_c)\simeq d$ with $R_c=At_c^{1/8}$. 
Figure~\ref{fig:hocking} shows that a nice collapse of drop radius versus time can be obtained for different particle sizes and drop volumes by normalising the radius by the critical radius $R_c$ and the time by the critical time $t_c$. 
Spreading deviates from \textcolor{black}{the power} law (solid line) for $t\simeq t_c$ for the 250 and \SI{550}{\micro\meter} particles when $R\simeq R_c$.
\textcolor{black}{This slowing down of the dynamics is accompanied by a change of convexity of the liquid/gas interface as seen in figure~\ref{fig:breakdown}($a$). This change may stem for various origins such as the constraint imposed by the edge of the particle layer on the interface height or the ability of the capillary stress driving the motion of the contact line to drain liquid out of the particle layer.}
For the smallest particle diameter, the experiments are too short to observe the deviation as the critical thickness $h_0=d$ is far from being reached.

\section{Concluding remarks}
\label{sec:conc}
A global-scale investigation of the spreading of a drop of granular suspension has been performed in the dense regime ($\phi\geq40\%$) in order to exhibit the strongest effects of the particle phase. 
The first major output, developed in \S~\ref{sec:effective_viscosity}, is that \textcolor{black}{the power law which describes the growth of the drop radius for large drops (i.e. when gravity prevails over capillarity)} is still valid provided one uses an effective viscosity independent of the particle size. 
This so-called \textcolor{black}{spreading} viscosity is much smaller than the bulk viscosity of the suspension probably due to the combined effect of particle slip near the tip of the drop and porous-like behaviour at the centre of the drop as shown by flow visualisation and PIV analysis in \S~\ref{sec:PIV}. 
The second important finding is that, when the height of the drop becomes of the order of the particle spacing, the spreading slows down.
This departure from \textcolor{black}{the power} law comes from particle freezing and drainage of the pure fluid out of the porous particle matrix.
In \S~\ref{sec:breakdown}, simply stating that this transition happens when the drop thickness reaches a particle
diameter, $h_0(t_c)=d$, provides a decent prediction of the breakdown of \textcolor{black}{the power} law at a solid volume fraction $\phi=40\%$.
However, a $\phi$-dependent criterion (which is difficult to measure) is expected in view of the earlier particle freezing seen at larger $\phi$ in figure~\ref{fig:breakdown}($b$), i.e. as $\phi$ becomes closer to a jamming volume fraction upon spreading, $\phi_c^s$, which is certainly lower than the usual bulk jamming volume fraction because of confinement \citep{scott1969density,desmond2009random}. 
We anticipate a weak algebraic divergence with the distance to the jamming point similar to that found for the transition neck diameter in the pinch-off of a viscous suspension thread \citep{chateau2018pinch}.
A tentative criterion could then be $h_0(t_c)=d/(\phi_c^s-\phi)^\alpha$ with $\alpha \simeq 1/3$ and could help to rationalise the earlier freezing when $\phi$ is close to $\phi_c^s$.
\textcolor{black}{The final stages of spreading of a granular drop and the transition from a continuous effective fluid to a discrete system are indicative of confinement-dependent jamming.}

\backsection[Supplementary data]{\label{SupMat}Supplementary material and movies are available at https://doi.org/10.1017/jfm.2023...}

\backsection[Acknowledgements]{We would like to thank two interns, Ma\"el Lebon and Siham Mekelleche, for their precious help at the beginning of this project. We would also like to thank Tom Witten for insightful discussions and beneficial comments on this project.}


\backsection[Declaration of interests]{The authors report no conflict of interest.}


\backsection[Author ORCIDs]{\\
Alice Pelosse http://orcid.org/0000-0002-4554-0604;\\
\'Elisabeth Guazzelli http://orcid.org/0000-0003-3019-462X;\\
Matthieu Roch\'e http://orcid.org/0000-0002-8293-7029.
}


\bibliographystyle{jfm}
\bibliography{biblio_tanner}
\end{document}